\documentclass{article}
\usepackage[utf8]{inputenc} 
\usepackage[T1]{fontenc}    
\usepackage{hyperref}       
\usepackage{url}            
\usepackage{booktabs}       
\usepackage{amsfonts}       
\usepackage{nicefrac}       
\usepackage{microtype}      
\usepackage[final,nonatbib]{neurips_2019}
\usepackage{multirow}
\usepackage{gensymb}
\usepackage[toc,page]{appendix}
\usepackage{graphicx}

\title{
High Resolution Forecasting of Heat Waves impacts on Leaf Area Index by Multiscale Multitemporal Deep Learning}
\author{%
Andrea Gobbi$^*$, Marco Cristoforetti$^*$, Giuseppe Jurman, Cesare Furlanello \\
Fondazione Bruno Kessler\\
Trento, Italy\\
\texttt{\{agobbi,mcristofo,jurman,furlan\}@fbk.eu} \\
$^*$ joint first author
}

\begin{document}
\maketitle

\begin{abstract}

Climate change impacts could cause progressive decrease of crop quality and yield, up to harvest failures. In particular, heat waves and other climate extremes can lead to localized food shortages and even threaten food security of communities worldwide. In this study, we apply a deep learning architecture for high resolution forecasting (300 m, 10 days) of the Leaf Area Index (LAI), whose dynamics has been widely used to model the growth phase of crops and impact of heat waves. LAI models can be computed at $0.1^\circ$ spatial resolution with an auto regressive component adjusted with weather conditions, validated with remote sensing measurements. However model actionability is poor in regions of varying terrain morphology at this scale (about 8 km at the Alps latitude). Our deep learning model aims instead at forecasting LAI by training multiscale multitemporal (MSMT) data from the Copernicus Global Land Service (CGLS) project for all Europe at 300m resolution and medium-resolution historical weather data. Further, the deep learning model inputs integrate high-resolution land surface features, known to improve forecasts of agricultural productivity. The historical weather data are then replaced with forecast values to predict LAI values at 10 day horizon on Europe. We propose the MSMT model to develop a high resolution crop-specific warning system for mitigating damage due to heat waves and other extreme events.
\end{abstract}

%
%
%
\section{Intro}
The growing concern for heat waves impacts on human and animal health is matched by the urgent need of mitigating long and short term damage to crops at all latitudes worldwide. The Leaf Area Index (LAI) has been widely used in modeling the growth phase of crops (\cite{ewert2004modeling,heuvelink2005effect,malone2002relationship,eik1966leaf}) and for monitoring the impact of the heatwaves on vegetation (\cite{albergel2019monitoring}). Notably, besides decreasing yield, heat waves impacts are relevant modifiers of phenology phases and then of crop quality, thus requiring to a timely implementation with mitigation procedures, ideally based on forecasts valid at the cultivar scale. Computation of LAI is usually based on an autoregressive component adjusted with weather conditions. The LDAS-Monde platform has recently shown the potential of assimilating 
enhanced estimates of land surface conditions and remote sensing measurements at medium resolution scales ($0.1^\circ$)   (\cite{jarlan2007analysis,albergel2019monitoring}).  Unfortunately, such scale is unusable in case of highly variable terrain morphology, e.g. in the Alps and far for practical actionability in general. 

The availability of LAI as a product of the Copernicus Global Land Service (CGLS) for all Europe at 300m resolution has opened a novel option for downscaling LAI at more actionable spatial scales. The CGLS PROBA-V’s wide view and polar orbit provide a revisit of every spot on Earth’s land every two days, thus enabling to build a new global composite every 10 days. In this context, we have developed MSMT, a Multi Scale Multi Temporal neural network architecture that estimates the CGLS LAI maps as a novel forecast product. The network is fed with medium-resolution historical weather data integrated with high-resolution Land Data conditions as extracted by a Geographical Analyis System (abbreviated here as GIS features). The MSMT model can compute a daily LAI raster at high-resolution for all the Europe. We then replace the historical weather data with medium-resolution forecast values in order to obtain a high resolution LAI forecast at European scale. Although in an initial phase, this research work opens the potential for developing high resolution predictive maps of LAI at world scale, thus enabling the implementation of warning system in agriculture to mitigate damage due to extreme events, such as heat waves.

\section{Data}
We used LAI maps from the CGLS PROBA-V (vegetation) mission for all Europe at 300m resolution, corresponding to about 200Mb every 10 days. Weather data (4 measures per day of temperature, pressure and precipitation) are from the  European Centre for Medium-Range Weather Forecasts (ECMWF) ERA5-Land database at a $0.1^\circ$ resolution for five seasons (2014-2018), all in the vegetation periods beginning of March to end of October. Inputs include a set of 12 raster data layers defining GIS features, as described in Tab.\ref{tab:01} (Appendix). All the data sets have been aligned and harmonized both in time and space in the bounding box $-14^\circ W\ 39^\circ E,\ 34^\circ S\ 74^\circ N$. Daily predictions on pixels for which LAI or weather inputs are missing have been filtered out in this study. In summary, one observation per day with spatial resolution equal to 300m has been produced for 25-50 Mln of valid data points per year, depending on the considered season (about 1.2 TB in total).

\begin{figure}[th!]
\includegraphics[width=\linewidth]{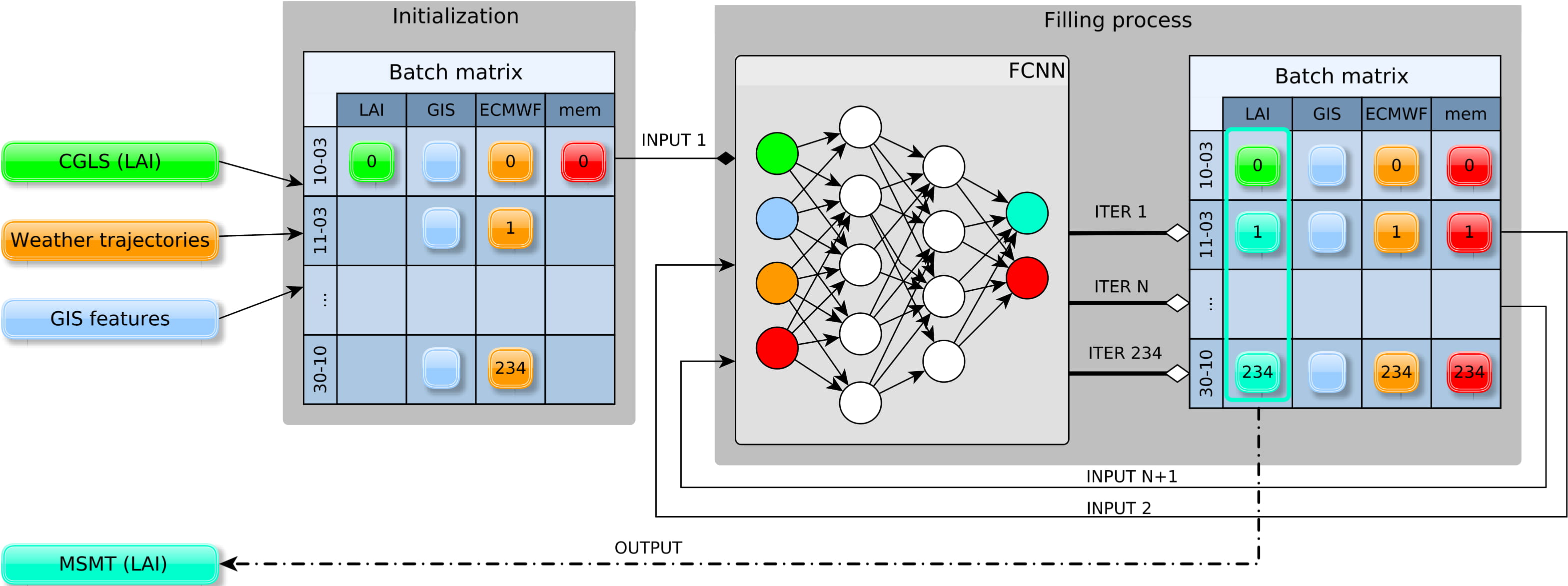}	
\caption{Model architecture}\label{fig:irnn}
\end{figure}

\section{The MSTM Model}
The MSMT model in this study computes for each pixel in Europe ($300m$ resolution) a LAI value for each day of the season. For each pixel, the model is initialized with the first LAI observation (around the 10th of March), the GIS features and the trajectories of the weather variables during the season, as sketched in Fig.~\ref{fig:irnn}. 
 Inspired to a recursive neural network architecture, the output of the MSMT network at time $t$ is part of the input at time $t+1$. In particular, the output of the network is composed by nine values: the predicted LAI at time $t$ and 8 values representing a \textit{memory} vector in which the network can buffer information about the time evolution of the LAI and weather data. The algorithm aims to capture some well known mechanisms of vegetation growth such as accumulation of temperatures and precipitations.
After seasonal initialization, the network is rerun for each day until the end of October. The result is the daily prediction for the LAI: since CGLS data are available each 10 days, only data in such timestamps are used to compute the loss and back propagate the error. 
Two embedding layers have been used to convert categorical variables (land use and soil) into numerical variables. After that, all the features (GIS, weather plus the memory features described above) pass through a fully connected neural network producing a the following value of the LAI and a new memory state. The network is then used in inference with the ERA-5 data to obtain forecasts. 
\section{Experiment and Results}
\begin{figure}[h!]
\includegraphics[width=\linewidth]{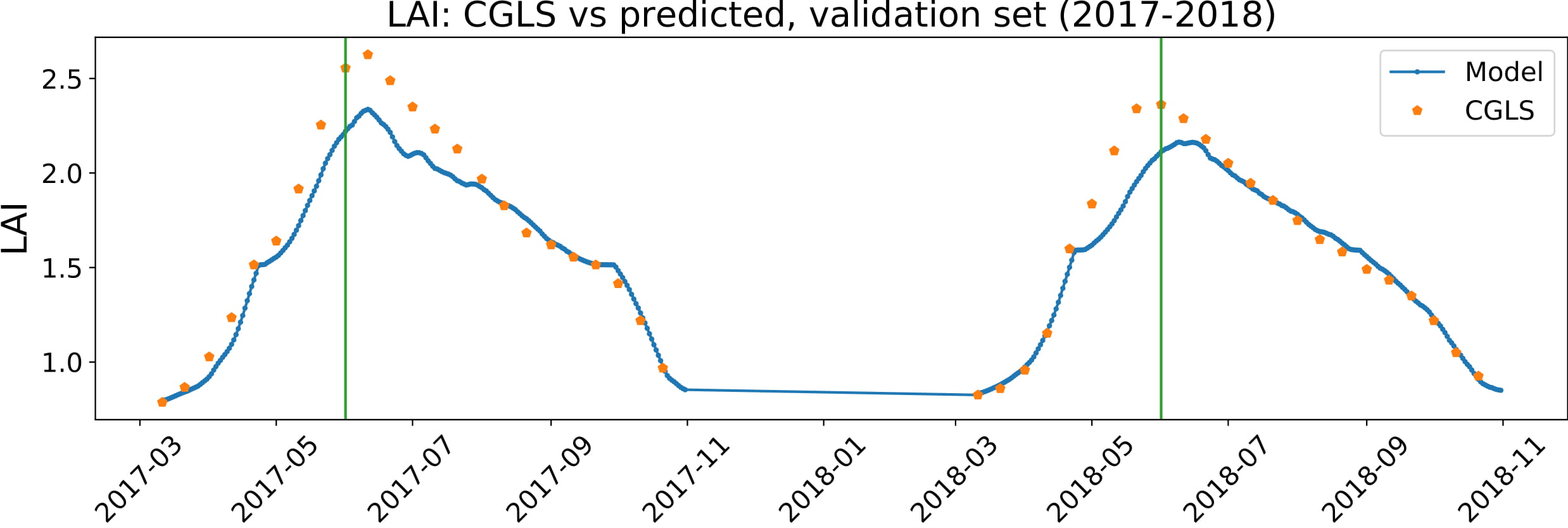}\label{fig:prediction}
\caption{ MSMT predictions for the validation set compared to LAI CGLS data (mean values). Orange dots:CGLS; blue line:MSMT. June 2017 and 2018 are highlighted in green.}
\end{figure}

Several configurations of MSMT hyperparameters were  tested (not reported here for lack of space) with experiments deployed in a CPU-based Azure VM with 64CPUs with 126Gb of RAM. 
Three seasons (2014, 2015 and 2016) were used as development sets while seasons 2017-18 were held out as external validation. The best model reaches RMSE = 0.738 (0.756) in season 2017 (2018), comparable with the results obtained by \cite{albergel2019monitoring}. In this model the two embedding layers have output size equal to 8 as the dimension of the memory state. We used a simple fully connected neural network (FCNN) with 3 hidden layer (64-128-64); the output size is 9 (1 LAI + 8 memory states).
The model is accurate along most of the season, especially  in the second part, even if it is based only on the LAI value at 10th of March and on the weather trajectories (Fig.\ref{fig:prediction}). In particular, MSMT correctly manages the heatwaves occurring during July 2018 producing accurate prediction during and after this period of interest. 

\begin{figure}[h!]
\centering
\includegraphics[width=\linewidth]{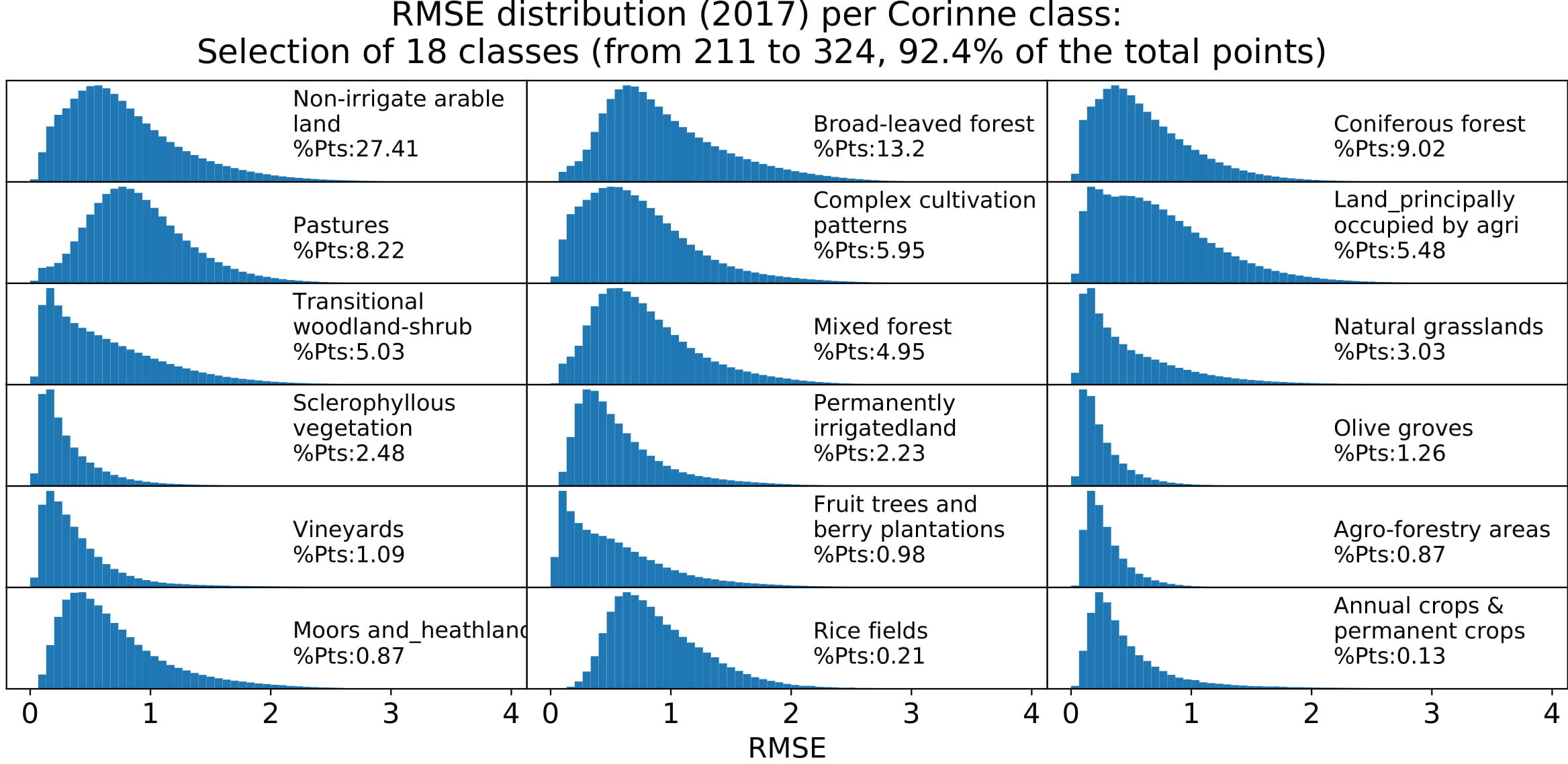}
\caption{RMSE (for season 2017) distribution over 18 selected land use classes. Names and percentage of coverage are reported inside each box.}\label{fig:corinne}
\end{figure}

\begin{figure}
\includegraphics[width=\linewidth]{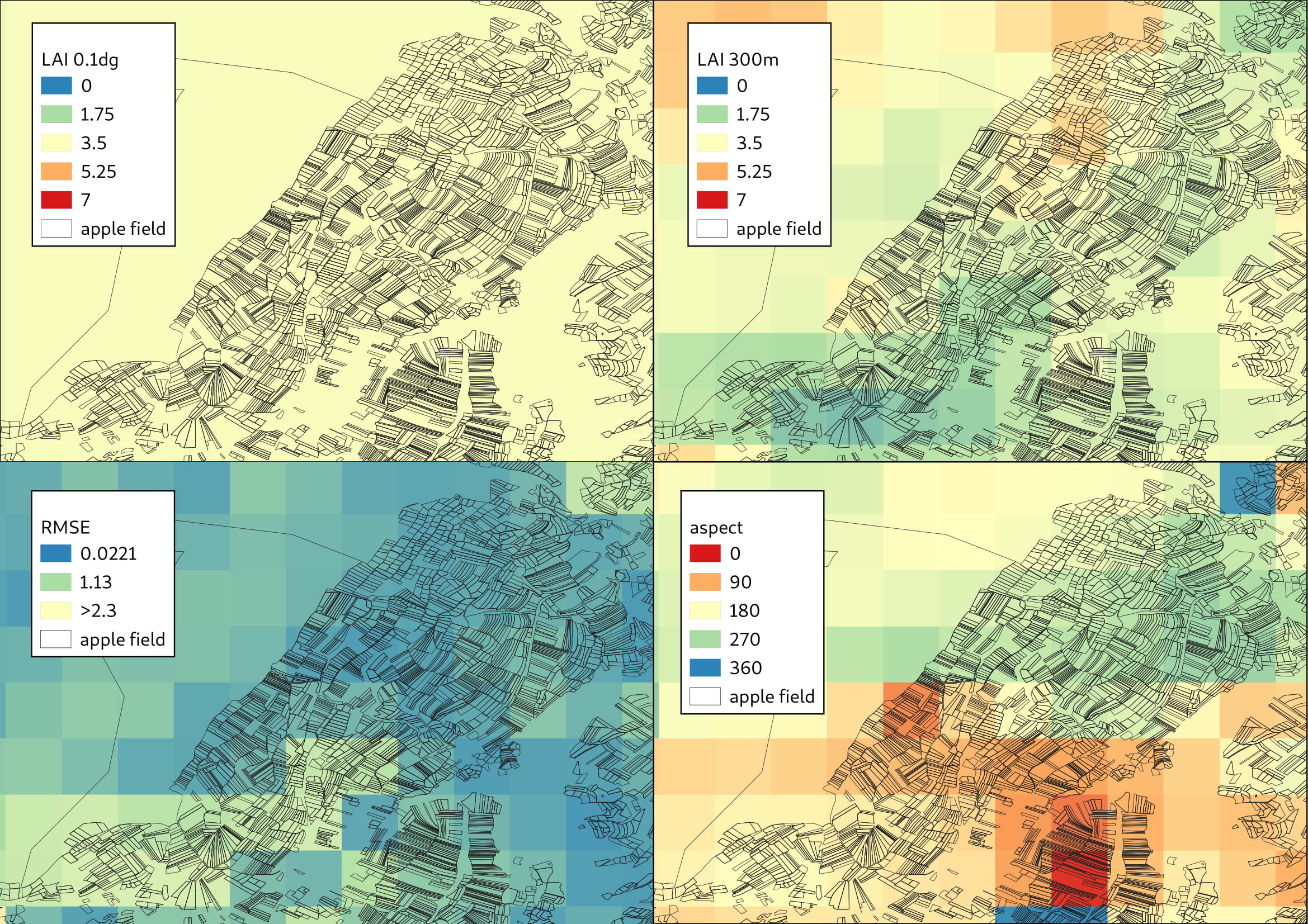}	
\caption{Analysis of LAI variability in a mountainous area. Polygons are apple fields. Top left: LAI value at medium resolution (ECMFW $0.1^\circ$) for the 10th August 2018. Top right: LAI at high resolution (CGSL 300m). Bottom left: RMSE error for the MSTM for the season 2018. Bottom right: aspect map (N: 90; E: 0=360) computed with GRASS GIS \texttt{r.slope.aspect} function.}\label{fig:maps}
\end{figure}

\begin{figure}[h!]
\includegraphics[width=\linewidth]{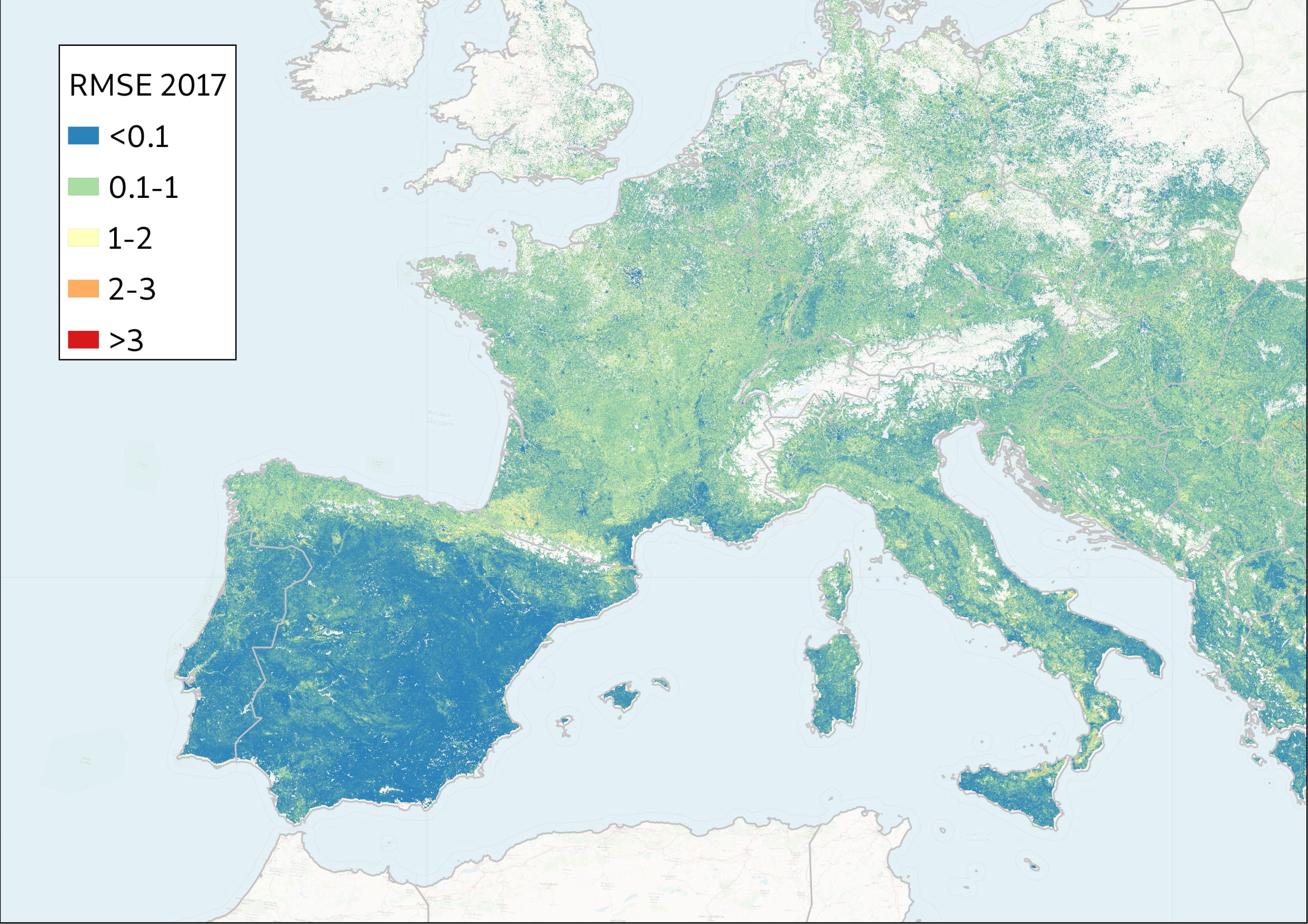}	
\includegraphics[width=\linewidth]{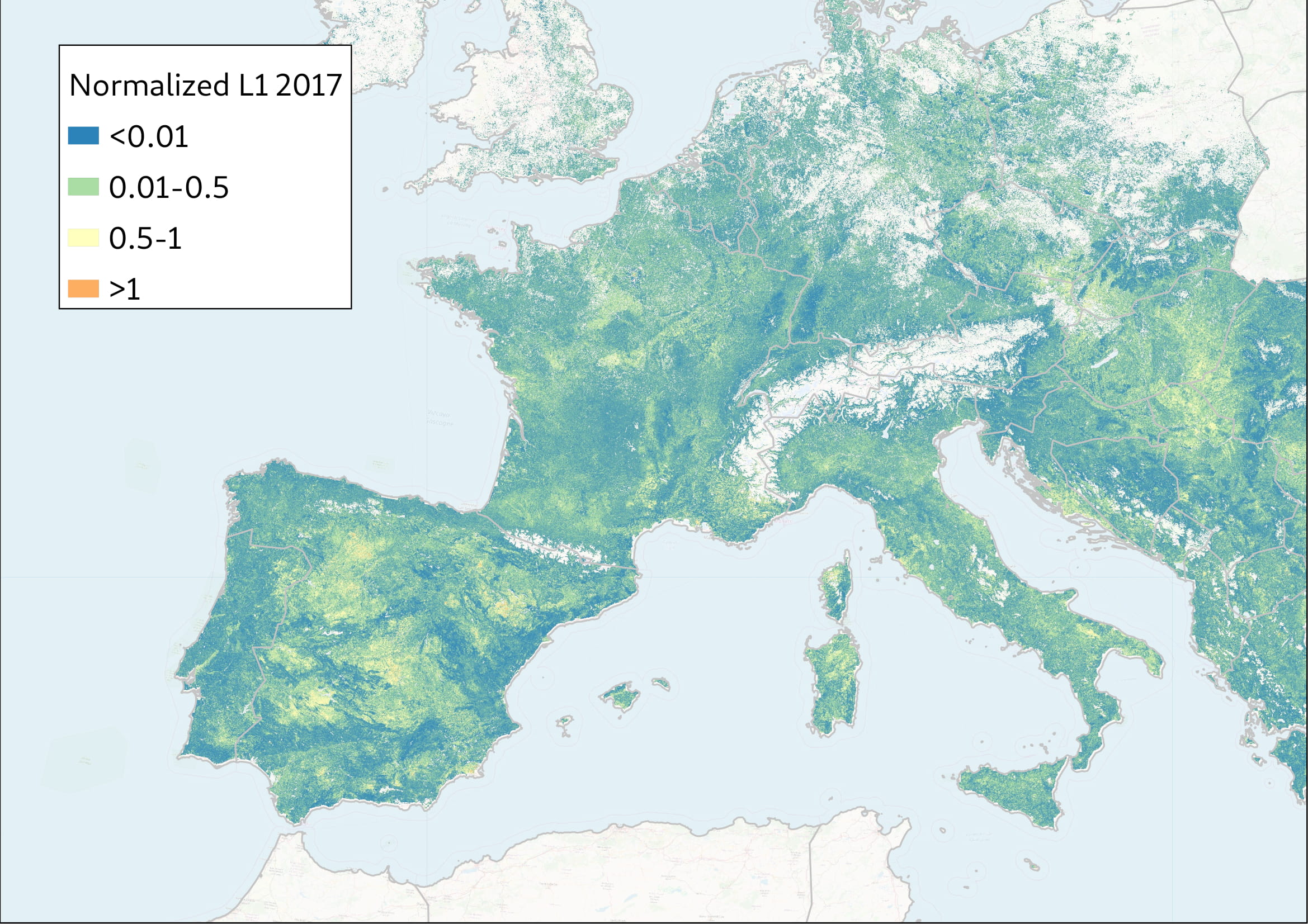}
\caption{MSMT Error maps (zoom on South-West Europe) for the season 2017. Top  RMSE, bottom: normalized  L1. Normalized L1 (pixel-wise operation: L1 divided by the mean value of the LAI) captures the variation of the prediction respect of the real values in terms of magnitude respect to absolute values. }\label{fig:sw}
\end{figure}

Land characterization (e.g. Corine class) seems to be  significant for the model error. The distribution of the RMSE over Corine classes are shown in Fig.\ref{fig:corinne} (Appendix). We are experimenting how to control land usage class and other GIS features to improve the results of the model, e.g. by excluding specific terrain patterns or weighting more others. This is relevant to address climate change impacts, in particular to optimize risk models such as the crop-specific LAI estimates. Examples of LAI maps, comparing state of art models and MSMT at 300m res, model error and errors with respect to CGLS are reported in Fig.\ref{fig:maps} (Appendix). Error maps for South-West Europe are also reported in Fig.\ref{fig:sw} (Appendix). 

\section{Conclusions}
We have introduce the MSMT deep learning architecture to obtain a novel forecast LAI map at the same high resolution (300 m) of the Copernicus product. This model is applicable at global scale, here demonstrated for whole Europe in combination with land surface conditions (GIS features), whose assimilation is being investigated to improve the model. Notably the MSMT network, coupled with a state of art weather forecasting system can be used to implement warning services in agriculture. In particular, we aim to support strategies to mitigate damage to crops during extreme events such as heatwaves. Future works will include experiments with other core networks and the development of crop specific prediction based on the LAI. For the fine tuning of the model we will explore training strategies dependent of Corine over land use classes and terrain morphology patterns.

\subsubsection*{Acknowledgments}
This work has been supported by the grant 'Microsoft AI for Earth'' and by the ''Partenariato Europeo per l’Innovazione'' project ''Clima ed Agricoltura in area montana" .

\clearpage
\bibliographystyle{unsrt}
\bibliography{gobbi2019high}

\clearpage

\newpage
\section*{Appendix}

\begin{table}[h!]
\tiny
  \caption{Resolution, source, type and frequency of input features. Part of GIS features has been processed with GRASS 7.6.}
  \label{tab:01}
  \centering
\begin{tabular}{p{2cm}p{0.8cm}p{7cm}p{1cm}p{1.2cm}}
  Name & Res. &Source & Type & Freq.\\
    \midrule
 Corine land use     & 25m     & \url{https://land.copernicus.eu/pan-european/corine-land-cover} & 44 classes & \multirow{10}{*}{static} \\
   Soil WRB     & 250m & \url{https://data.isric.org/geonetwork/srv/eng/catalog.search#/metadata/5c301e97-9662-4f77-aa2d-48facd3c9e14}  & 118 classes &  \\
    DTM     & 25m     & \url{https://land.copernicus.eu/imagery-in-situ/eu-dem/eu-dem-v1.1/view}& \multirow{6}{*}{float}  &\\
    lat, lon     & 25m     & DTM  & & \\
    pcurv*, tcurv*, slope* and aspect* & 25m & DTM + \textbf{r.slope.aspect} default parameters &  & \\
    relief* & 25m & DTM + \textbf{r.relief} default parameters & & \\
    strength* and fisher* & 25m & DTM + \textbf{r.roughness.vector} windows size = 7 &   &\\
    \bottomrule
 min/max temp., pressure, sum/avg precip. & $0.1^\circ$ & \url{https://cds.climate.copernicus.eu/cdsapp#!/dataset/reanalysis-era5-land?tab=overview} &float& 4 per day \\   
    \bottomrule
    LAI CGLS & 300m & \url{https://land.copernicus.eu/global/products/lai} & float& every 10 days \\
      \bottomrule
  \end{tabular}
\end{table}


\end{document}